# Stress-induced R-$M_A$-$M_C$-T symmetry changes in $BiFeO_3$ films

J.H. Nam[1,2], H.S. Kim[1], A. J. Hatt[3], N. A. Spaldin[3], and H.M. Christen[1]

[1]Materials Science and Technology Division, Oak Ridge National Laboratory, Oak Ridge, TN 37831, USA
[2]Optic and Electronic Ceramics Division, Korea Institute of Ceramic Engineering and Technology (KICET), Seoul 153-801, Republic of Korea
[3]Materials Department, University of California, Santa Barbara, CA 93106, USA

The recent discovery of a stress-induced structural phase transition in the single-component perovskite $BiFeO_3$ (refs 1-3) has revived interest in this lead-free ferroelectric and points to new applications of a material already being the most promising multiferroic[4]. The coexistence of different phases may lead to large piezoelectric coefficients[3], a property that is typically associated with complex solid solutions of lead-based perovskites. In this work, we report combined experimental and computational results showing that the stress-induced phase transitions in $BiFeO_3$ follow the path of rhombohedral(R)-to-monoclinic($M_A$)-to-monoclinic($M_C$)-to-tetragonal(T), where both $M_C$ and T show highly enhanced *c/a* ratios of out-of-plane (*c*) to in-plane (*a*) lattice parameters. This R-$M_A$-$M_C$-T sequence is otherwise observed only near morphotropic phase boundaries (MPBs) in lead-based perovskites (i.e. near a compositionally induced phase instability), where it is controlled by electric field, temperature, or composition (refs 5-7). Our results represent the first time that this evolution has been induced in a single component system using strain alone, and show that substrate- imposed symmetry lowering results in a similar phase instability as the proximity to a MPB in solid solutions.

Epitaxial films provide an ideal platform to study the effect of biaxial stress on complex materials. When the rhombohedral perovskite $BiFeO_3$ is grown on a substrate having a square in-plane lattice, its symmetry is lowered to monoclinic, except in free-standing membranes and very thick films that remain rhombohedral. At tensile to moderate-compressive strains, the monoclinic structure loosely resembles that of the rhombohedral bulk, and we refer to it as the "R-like" phase. Compressive strains exceeding about -4.5% result in a strongly increased out-of-plane lattice parameter $c_{pc}$, leading to a $c_{pc}/a_{pc}$ ratio of about 1.25 (where $a_{pc}$ is the in-plane lattice parameter and the subscript *pc* refers to the pseudo-cubic notation) and an abrupt change in Fe coordination. This monoclinic phase thus resembles that of a super-tetragonal perovskite, and we therefore refer to it as "T-like".

Both "R-like" and "T-like" phases are ferroelectric. The polarization ***P*** points approximately along the $[111]_{pc}$ direction in the "R-like" phase[8,9] and rotates towards the $[001]_{pc}$ direction with increasing compressive strain. For the "T-like" phase, the very large values of the measured projection of ***P*** onto the $[001]_{pc}$ direction[1,2] agree with theoretical predictions of a polarization predominantly along $[001]_{pc}$ (ref 10), although surprisingly there is a report of a strong in-plane component using piezoresponse force microscopy[11].

The transition between the two monoclinic "R-like" and "T-like" phases has been labeled *iso-symmetric*; in this work we show that this is an oversimplification and in fact the evolution with increasing strain is from an "R-like" monoclinic $M_A$ phase of *Cc* or *Cm* symmetry to a "T-like" monoclinic $M_C$ phase of *Pc* or *Pm* symmetry with increasing strain. Then at very high strain values, which have not yet been accessed through direct coherent heteroepitaxy alone, a fully tetragonal *P4mm* phase is obtained.

First we describe the structures of the relevant phases. Consistent with all literature and current experimental and computational results we take $\beta \neq 90°$ to be between the *a* and *c* axes. Since only four monoclinic space groups (*Pm, Pc, Cm*, and *Cc*) can show a non-vanishing polarization in the *c* direction, we restrict our discussion to the corresponding primitive (*P*) or base-centered (*C*) unit cells containing either a mirror (*m*) or glide (*c*) plane perpendicular to the *b* axis; ***P*** is then constrained by symmetry to its mirror or glide plane, ***P*** ‖ [$u0v$]. We use standard notation[12] to distinguish three monoclinic cases based on the orientation of ***P*** with respect to the pseudocubic coordinates: $M_A$ and $M_B$ for ***P*** ‖ [$uuv$]$_{pc}$, with $u<v$ and $u>v$, respectively; and $M_C$ for ***P*** ‖ [$u0v$]$_{pc}$ . (The tetragonal (T), rhombohedral (R), and orthorhombic (O) phases correspond to a constraint of ***P*** to a *symmetry axis* along [001]$_{pc}$, [111]$_{pc}$, or [011]$_{pc}$, respectively, see Fig. 1a).

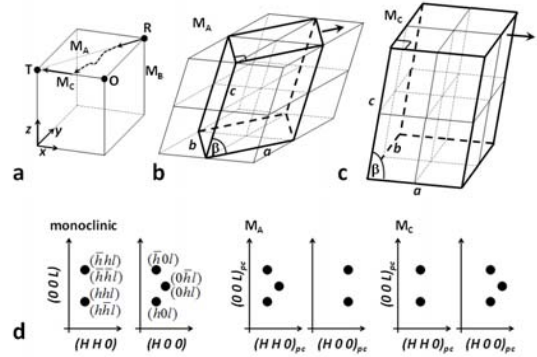

**Figure 1.** Schematic comparison of monoclinic structures. **a**: polarization direction in differently distorted perovskite structures: the polarization points in a direction parallel to that connecting the origin with the indicated dots for R, O, and T, or with any point on the lines labeled $M_A$, $M_B$, and $M_C$. The path R – $M_A$ – $M_C$ – T observed here and in relaxor ferroelectrics is indicated by arrows (see text). **b** and **c**: Smallest unit cells (with appropriate antiferromagnetic order and octahedral tilt patterns) for the two possible monoclinic structures, each resulting from a shear distortion in the direction of the indicated arrow. Thin lines indicate the primitive perovskite unit cell. $M_A$ is a centered unit cell belonging to space group *Cc* or *Cm*, $M_C$ is a primitive unit cell, space group *Pc* or *Pm*. **d**: Schematic representation of reciprocal space maps of a monoclinic structure, indexed in the monoclinic system and the pseudocubic notation for $M_A$ and $M_B$, respectively.

$M_A$ and $M_B$ both correspond to either *Cm* or *Cc* space groups, with a monoclinic unit cell rotated by 45° with respect to the pseudocubic structure (i.e. [$u0v$] ‖ [$uuv$]$_{pc}$), and are obtained by shearing the cubic perovskite cell along the [110]$_{pc}$ direction (Fig. 1b). For simplicity we restrict our discussion to $M_A$. In contrast, $M_C$ corresponds to either *Pm* or *Pc*, resulting from a shear along [100]$_{pc}$. In Figs. 1b and c we sketch the smallest $M_A$ and $M_C$ unit cells that are also compatible with common types of antiferromagnetism and octahedral tilt patterns[13]. This requires a √2 x √2 x 2 supercell for the (centered) $M_A$ and a 2 x 2 x 2 supercell for the (primitive) $M_C$ phase.

We study the structure of BiFeO$_3$ at different compressive biaxial strains in epitaxial films on SrTiO$_3$ (001), LaAlO$_3$ (001)$_{pc}$ and YAlO$_3$ (001)$_{pc}$ substrates. All samples are single-phase, epitaxial, with only the (00$l$)$_{pc}$ peaks detected in normal θ-2θ x-ray diffraction (XRD) scans. 270 nm thick films on SrTiO$_3$ substrates show a $c_{pc}$-axis lattice parameter of 3.99Å, corresponding to the "R-like" phase, as in our previous work[8], while for those on LaAlO$_3$ we determine $c_{pc}$ = 4.64Å, i.e. we find the "T-like"-phase (again in films with a thickness of 270 nm). Growth of monoclinic BiFeO$_3$ on a higher-symmetry substrate requires the formation of a multi-domain structure. We observe single peaks in the ω-scans (rocking curves) through the (00$l$)$_{pc}$ film positions (data not shown) showing that the (00$l$)$_{pc}$ planes of all domains are parallel to each other, as also reported elsewhere[14-16]. Reciprocal space maps (RSMs) through the {113}$_{pc}$ and the {103}$_{pc}$ family of peaks are shown in Fig. 2a. The observation of a two-fold and three-fold splitting along the [113]$_{pc}$ and the [103]$_{pc}$ directions, respectively, indicates the type of monoclinic distortions. In fact, as illustrated in Fig. 1d, it is easily seen that the monoclinic family of peaks splits three-fold for {$hhl$} and two-fold for {$h0l$} (refs17-18). With the 45° rotation of $a$ with respect to $a_{pc}$ in the M$_A$ monoclinic structure but $a \parallel a_{pc}$ in M$_C$, the RSMs for the two monoclinic structures look distinctively different. Clearly, our results for "T-like"-BiFeO$_3$/LaAlO$_3$ are consistent only with the M$_C$ structure, and thus distinctively different from that in "R-like"-BiFeO$_3$, for which our data shown below and that of a number of authors[8,14-19] clearly indicate the M$_A$ structure.

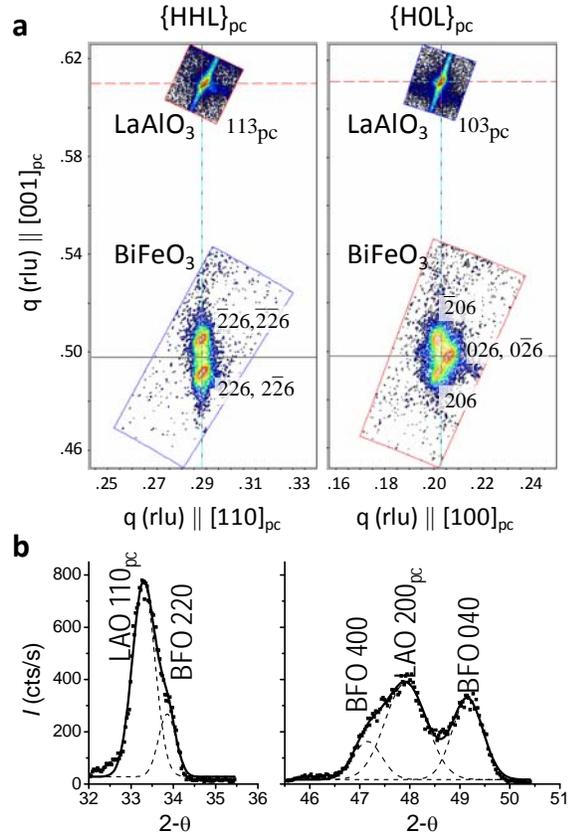

**Figure 2.** X-ray diffraction data for "T-like"-BiFeO$_3$ on LaAlO$_3$. **a**: RSMs in the pseudocubic [113] and [103] directions. **b**: In-plane diffraction θ-2θ scans through the substrate's 110$_{pc}$ and 200$_{pc}$ peaks.

The M$_C$ symmetry of the "T-like" phase is confirmed in in-plane (grazing-incidence) XRD scans through the substrate's 110$_{pc}$ and 100$_{pc}$ peaks (Fig. 2b), which shows single peak along the pseudocubic face diagonal, but a split peak along 100$_{pc}$. From a refinement of 20 diffraction peaks combined with relative (peak separation) measurements from these RSMs and in-plane scans we find for "T-like"-BiFeO$_3$ $a$ = 2 x 3.84(2) Å, $b$ = 2 x 3.70(2) Å, $c$ = 2 x 4.64(2) Å, and $β$ = 87.9(2)° (where the doubling of the unit cell is not seen in the data but considered for compatibility with antiferromagnetism and octahedral tilts). Thus, these data for the first time show that monoclinic "T-like"-BiFeO$_3$ exhibits a M$_C$ structure belonging to either the *Pm* or *Pc* space groups. Note that Raman data[11], while originally interpreted in terms of *Cc*, are insufficient to determine the type of monoclinic distortion. However, earlier x-ray data presented but not fully analyzed by Iliev *et al.*[20] are fully consistent with our observations.

Previous first principles calculations found the $M_A$ (*Cc*) structure to be the ground state for "T-like"-BiFeO$_3$ (ref 10) but an exhaustive search of all possible monoclinic shear orientations, polarization orientations, c/b ratios and monoclinic angles, was not previously performed. Therefore, here we make such a search, looking explicitly for the $M_C$ phase, and using a 2 x 2 x 2, 40 atom unit cell that accommodates the distortions expected for the $M_C$ phase. We find that G-type antiferromagnetic order is energetically indistinguishable from C-type for a few representative structures, and hereafter maintain G-type for all calculations.

To most closely simulate the experimental conditions, we constrain the in-plane lattice parameters to the measured values scaled by the ratio of experimental to local spin density approximation (LSDA) values of the bulk lattice parameter, (giving *b* = 2 x 3.69 Å), and we use the experimental *c/b* = 1.25. We then search for the lowest energy structure for shear orientations along $[110]_{pc}$ and $[100]_{pc}$, corresponding to $M_A$ and $M_C$, respectively, relaxing *β* and *a/b* for each. In our search for the ground state, we initialize the system to a number of likely polarization orientations and patterns of octahedral rotations to explore the structural phase space as completely as possible within the computational time limitations.

We find three phases that are very close in energy: the $M_A$ (*Cc*) phase studied in previous work remains lowest in energy[10], but we also identify two low energy metastable phases with shear along $[100]_{pc}$ corresponding to space groups *P1* and *Pm*. As before, the $M_A$ (*Cc*) phase has $a^-b^-c^0$ rotations and a polar displacement along the glide plane; in contrast, the $M_C$ (*Pm*) phase has $a^-b^0c^0$ rotations and a polar displacement along the mirror plane. The triclinic *P1* phase has $a^-b^-c^0$ rotations but a polar displacement with components along all three lattice directions. Additional structural details are summarized in Supplementary Table 1 and Supplementary Figure 1. Note that a continuous transition from $M_A$ to $M_C$ would require an intermediate phase (dashed arrow in Fig. 1a)[12,21], such as this *P1* structure. The energies of the metastable $M_C$ (*Pm*) and the *P1* phases are higher in energy than the $M_A$ phase by only ~4 meV per formula unit; other possible phases that have been considered in the literature, such as the Cm (ref 22) and tetragonal P4mm are considerably higher in energy (~15 and ~20 meV per formula unit respectively). Note that previously calculated domain wall energies in R-BiFeO$_3$ (~100 meV per unit cell)[23] are also two orders of magnitude larger than the energy differences between these various "T-like" phases, suggesting that factors such as the ability to form favorable domain patterns might be more important than absolute total energies of different phases in determining the ground state structure adopted by a strained film.

Next we explore whether at higher strain values the monoclinicity can be completely removed and a genuine tetragonal phase can be formed. When we impose in our calculations an exceedingly large in plane strain of 10% (*a* = *b* = 2 x 3.5 Å) we indeed find a tetragonal (*P4mm*) ground state with *c / a* = 1.41 and the octahedral rotations reduced to zero.

To verify this computational result experimentally we first grew films on orthorhombic (110)-oriented YAlO$_3$ substrates, having a pseudocubic in-plane lattice parameter of 3.704 Å (ref 24).

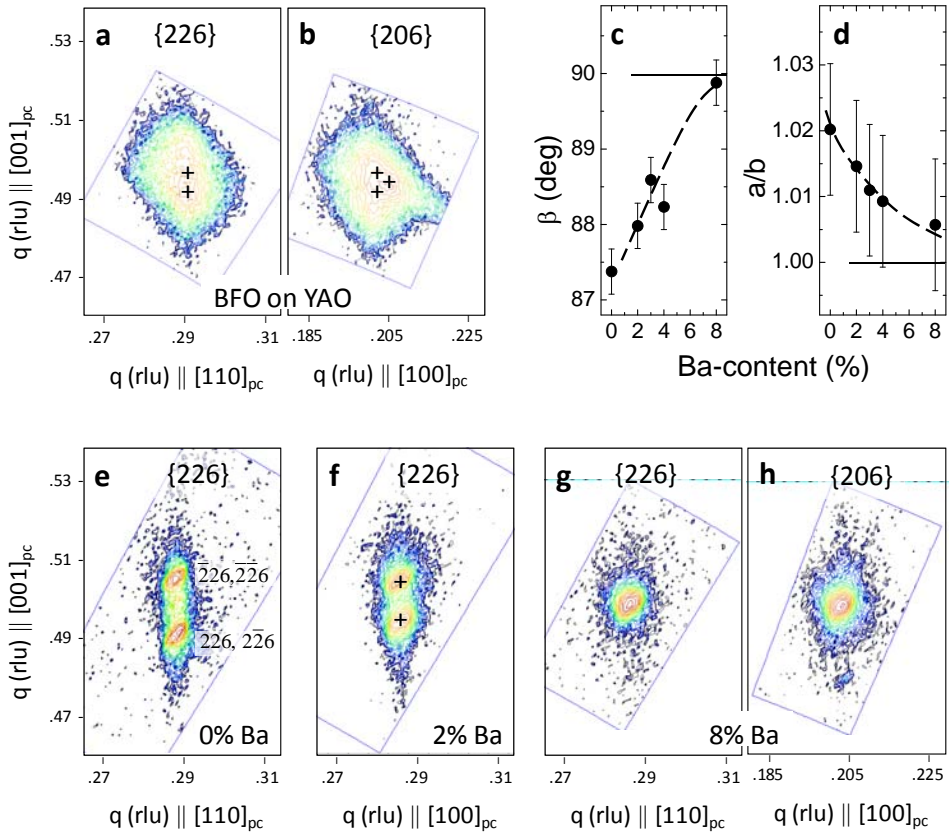

**Figure 3.** Strain and composition effects on the $M_C$ phase. **a-b**: RSMs through the {226} and {206} peaks (i.e. in the $[113]_{pc}$ and $[103]_{pc}$ directions, respectively) for BiFeO$_3$ grown on YAlO$_3$. Approximate peak positions are indicated by crosses. **c-d**: Evolution of the lattice parameters as function of Ba-content in Bi$_{1-x}$Ba$_x$FeO$_3$ on LaAlO$_3$. Dashed lines are guides to the eye. **e-h**: RSMs through the {226} peaks for different values if $x$, and the {206} peaks for $x = 0.08$.

Normal θ-2θ x-ray scans show that the BiFeO$_3$ separates both into the "R-like" and "T-like" phases (Supplementary Figure 2). Nevertheless, the peaks corresponding to the "T-like"-phase can again be indexed as $M_C$, but with $a = 2 \times 3.82(4)$ Å, $b = 2 \times 3.72(4)$ Å, $c = 2 \times 4.66(2)$ Å, and $\beta = 88.5(3)°$. Therefore, the $M_C$ symmetry is preserved but the structure becomes closer to tetragonal under additional stress, with $b/a$ closer to unity and $\beta$ closer to 90°. The RSMs for this sample are shown in Figure 3 a and b.

To induce a larger effective strain, we next modify our "T-like" BiFeO$_3$ via the substitution of Ba for Bi to make a Bi$_{1-x}$Ba$_x$FeO$_{3-\delta}$ solid-solution with $\delta \approx x/2$ (ref 25). As we have shown previously[25], this substitution results in an enlarged unit cell volume – as a consequence of both the larger ionic radius of Ba$^{2+}$ and the formation of oxygen vacancies – and therefore a larger effective biaxial compressive stress when grown on a same substrate. In Fig. 3e-g, we show RSMs through the {226} family $x = 0$, 0.02, and 0.08. A clear transition from the monoclinic ($M_C$) to a tetragonal structure is observed, with no measurable peak splitting in either {226} or

{206} (Fig. 3h) $x = 0.08$. Figure 3c and d show the evolution of $\beta$ and the ratio of the in-plane lattice parameters, $b/a$, as a function of Ba content, showing the gradual transition from the monoclinic to the tetragonal structure.

While the stabilization of this true tetragonal structure is not purely due to epitaxial strain, in combination with the data for BiFeO$_3$ on YAlO$_3$ and the computational results, it implies that additional biaxial stress applied to the M$_C$ structure will result in a true tetragonal phase. Therefore, BiFeO$_3$ can undergo the complete R-M$_A$-M$_C$-T path of transitions. This is best summarized by comparing RSMs for four different samples (see Fig. 4), corresponding each to a different symmetry. The 2 μm thick BiFeO$_3$ film on SrTiO$_3$ (bottom panel) is essentially relaxed due to its large thickness, and shows a diffraction pattern similar to that found for rhombohedral BiFeO$_3$ elsewhere[9,26]. A thinner film of BiFeO$_3$ shows the typical diffraction pattern for "R-like" BiFeO$_3$ with the M$_A$ structure as discussed elsewhere[8,14-19]. For the "T-like" BiFeO$_3$ sample (M$_C$ structure) and the Bi$_{0.92}$Ba$_{0.08}$FeO$_3$ (T) the data in Fig. 4 are taken from Figs. 2 and 3, respectively.

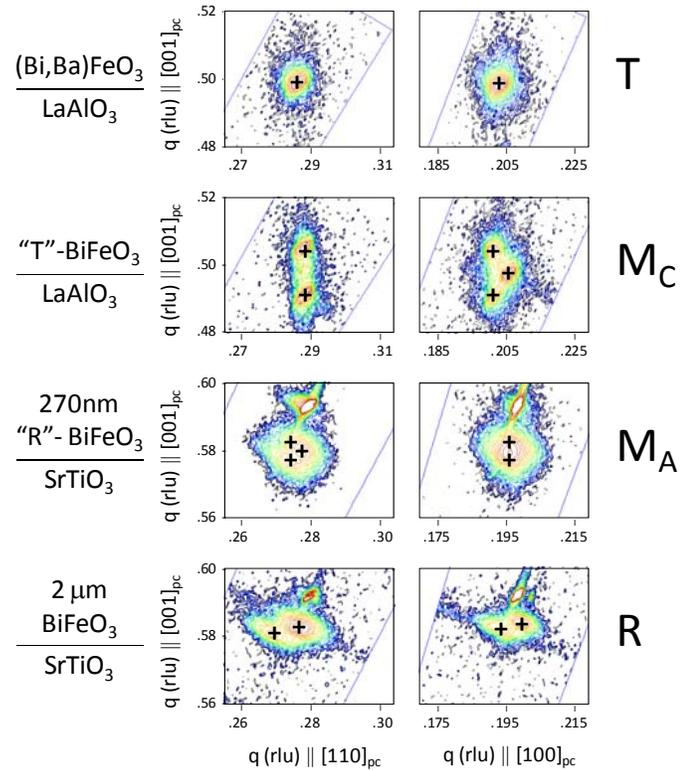

**Figure 4.** RSMs in the [113]$_{pc}$ (left column) and [103]$_{pc}$ (right) direction, for structures as indicated by the labels on the left. Effective stress increases from bottom to top. The path R-M$_A$-M$_C$-T corresponds to that indicated in Fig. 1a.

Symmetry changes as a consequence of external parameters have been observed in BiFeO$_3$ in a number of experiments. For example, films on SrTiO$_3$ show a transition near 750°C from M$_A$ to M$_C$ with temperature[27], presumed to be associated with the ferroelectric-to-paraelectric phase transition. Isostatic pressure applied to powders results in a transition from *R3c* to a non-ferroelectric *C2/m* monoclinic phase[28]. Computational results indicate a broad variety of phase changes with electric field[29]. However, the changes observed here are most similar in nature to those observed in the lead-based relaxor ferroelectrics (1-$x$)Pb(Mg$_{1/3}$Nb$_{2/3}$)O$_3$ – $x$(PbTiO$_3$) (PMN-PT) and (1-$x$)Pb(Zn$_{1/3}$Nb$_{2/3}$)O$_3$ – $x$(PbTiO$_3$) (PZN-PT). In fact, synchrotron x-ray diffraction data[5] on PZN-PT lead to a (field, composition)-phase diagram in which the R-M$_A$-M$_C$-T path can be traversed either in the field or composition direction. Similarly, the R-M$_A$-M$_C$-T path is also seen in the (field, temperature)-phase diagram of PMN-PT (refs 6-7) and PZN-PT (ref 7). These last results further suggest the possibility of similar but stress-induced structural changes, but there has been no experimental observation thereof. Intriguingly, calculations for Pb(Zr$_{1-x}$Ti$_x$)O$_3$ indicate the presence of a triclinic phase when the path from M$_A$ to M$_C$ is traversed via application of an electric field[21] or changes in composition[12]. Our computational results for BiFeO$_3$ show the energetic proximity of the M$_C$ and a triclinic (*P1*) phase and thus further illustrate the analogies between this single-component perovskite and the lead-oxide solid solutions.

In conclusion, the results presented here clarify that the "R-like"-to-"T-like" transition in BiFeO$_3$ is not as previously thought "iso-symmetric" but part of a broader R-M$_A$-M$_C$-T path in which the polarization orientation changes from one lying within the $(1\bar{1}0)_{pc}$ plane to one in the $(010)_{pc}$ plane. The observation of this R-M$_A$-M$_C$-T phase sequence in a stoichiometric compound, rather than in a solid-solution near a MPB, allows us to draw important conclusions regarding the control of structural and ferroelectric phase transformations and the potential of achieving technologically important piezoelectric properties in lead-free materials. In the lead-oxide solid solutions, the stability of the monoclinic phases is limited to a narrow region in composition space near the MPB. Given the delicate energy balance between the involved phases and possible local variations in stoichiometry, open questions remain in these solid solutions regarding the origins of these phase transitions and the mechanisms that can be used to control them (see ref 30 for a recent review). Our experimental and computational observations on BiFeO$_3$ now show that a stoichiometric material can mimic the behavior of a solid-solution near its MPB, illustrate the strong effect of substrate-induced symmetry lowering, and demonstrate that the R-M$_A$-M$_C$-T path can be controlled by epitaxial strain alone. This shows that "strain engineering" is an important tool both in fundamental studies to understand complex phase equilibria as well as in approaches to find new, lead-free materials with technologically relevant properties, such as large piezoelectric coefficients.

**Methods**

Epitaxial films were grown by pulsed-laser deposition at a substrate temperature of 700°C, a background pressure of 25 - 50 mTorr, using a KrF excimer laser (wavelength 248 nm, energy per pulse 550 - 590 mJ, pulse repetition rate 5 Hz) and sintered targets containing a 10% excess of Bi. At the target-to-substrate distance of 50 mm, the resulting deposition rate was ~0.2 Å/pulse. Solid-solution films (Bi$_{1-x}$Ba$_x$FeO$_3$) were obtained by a method in which submonolayer amounts of BiFeO$_3$ and BaFeO$_3$ are sequentially and repeatedly deposited by ablating from two separate targets[31]. All samples were then characterized using a PANalytical X'Pert Pro 4-circle x-ray diffractometer using Cu K$\alpha$ radiation. Reciprocal space maps (RSMs) were recorded by a sequence of $\theta$-2$\theta$ scans at different $\omega$ offsets, and are shown in plots of intensity as a function of $q$ in reciprocal lattice units (rlu), where $q = \pi/4d$.

First principles calculations were carried out using density functional theory as implemented in the Vienna Ab-initio Simulation Package, employing the PAW method and the local spin density approximation (LSDA) plus a Hubbard U term on the Fe d states, as described elsewhere [Hatt]. We achieve well-converged total energies using a 500 eV plane-wave cutoff and a 4x4x4 k-point grid. Space group determination was performed with the software FINDSYM (ref 32).

## References

1. Ricinschi, D., Yun, K.-Y., & Okuyama, M. A mechanism for the 150 μC cm$^{-2}$ polarization of BiFeO$_3$ films based on first-principles calculations and new structural data. *J. Phys.: Condens. Matter* **18**, L97 (2006).
2. Béa, H. *et al.* Evidence for Room-Temperature Multiferroicity in a Compound with a Giant Axial Ratio. *Phys. Rev. Lett.* **102**, 217603 (2009).
3. Zeches, R. J. *et al.* A Strain-Driven Morphotropic Phase Boundary in BiFeO$_3$. *Science* **326**, 977–980 (2009).
4. Wang, J. *et al.* Epitaxial BiFeO$_3$ Multiferroic Thin Film Heterostructures. *Science* **299**, 1719–1722 (2003).
5. Noheda, B. *et al.* Electric-field-induced phase transitions in rhombohedral Pb(Zn$_{1/3}$Nb$_{2/3}$)$_{1-x}$Ti$_x$O$_3$. *Phys. Rev. B* **65**, 224101 (2002).
6. Bai, F. *et al.* X-ray and neutron diffraction investigations of the structural phase transformation sequence under electric field in 0.7Pb(Mg$_{1/3}$Nb$_{2/3}$)-0.3PbTiO$_3$ crystal. *J. Appl. Phys.* **96**, 1620 (2004).
7. Davis, M., Damjanovic, D. & Setter, N. Electric-field-, temperature-, and stress-induced phase transitions in relaxor ferroelectric single crystals. *Phys. Rev. B* **73**, 014115 (2006).
8. Kim, D. H., Lee, H. N., Biegalski, M. D. & Christen, H. M. Effect of epitaxial strain on ferroelectric polarization in multiferroic BiFeO$_3$ films. *Appl. Phys. Lett.* **92**, 012911 (2008).
9. Jang, H. W. *et al.* Strain-Induced Polarization Rotation in Epitaxial (001) BiFeO$_3$ Thin Films. *Phys. Rev. Lett.* **101**, 107602 (2008).
10. Hatt, A. J., Spaldin, N. A. & Ederer, C. Strain-induced isosymmetric phase transition in BiFeO$_3$. *Phys. Rev. B* **81**, 054109 (2010).
11. Mazumdar, D. *et al.* Nanoscale Switching Characteristics of Nearly Tetragonal BiFeO$_3$ Thin Films. *Nano Lett.* **10**, 2555–2561 (2010).
12. Vanderbilt, D. & Cohen, M. H. Monoclinic and triclinic phases in higher-order Devonshire theory. *Phys. Rev. B* **63**, 094108 (2001).
13. Glazer, A. M. The classification of tilted octahedral in perovskites. *Acta Cryst*. **B28**, 3384 (1972).
14. Qi, X. *et al.* High-resolution x-ray diffraction and transmission electron microscopy of multiferroic BiFeO$_3$ films. *Appl. Phys. Lett.* **86**, 071913 (2005).
15. Daumont, C. J. M. *et al.* Tuning the atomic and domain structure of epitaxial films of multiferroic BiFeO$_3$. *Phys. Rev. B* **81**, 144115 (2010).
16. Liu, H., Yang, P., Yao, K. & Wang, J. Twinning rotation and ferroelectric behavior of epitaxial BiFeO$_3$ (001) thin film. *Appl. Phys. Lett.* **96**, 012901 (2010).
17. Saito, K. *et al.* Structural Characterization of BiFeO$_3$ Thin Films by Reciprocal Space Mapping. *Jpn. J. Appl. Phys.* **45**, 7311–7314 (2006).
18. Xu, G. *et al.* Low symmetry phase in (001) BiFeO$_3$ epitaxial constrained thin films. *Appl. Phys. Lett.* **86,** 182905 (2005).
19. Holcomb, M. B. *et al.* Probing the evolution of antiferromagnetism in multiferroics. *Phys. Rev. B* **81**, 134406 (2010).
20. Iliev, M. N., Abrashev, M. V., Mazumdar, D., Shelke, V. & Gupta, A. Polarized Raman spectroscopy of nearly tetragonal BiFeO$_3$ thin films. *Phys. Rev. B* **82**, 014107 (2010).
21. Bellaiche, L., García, A. & Vanderbilt, D. Electric-field induced polarization paths in Pb(Zr$_{1-x}$Ti$_x$)O$_3$ alloys. *Phys. Rev. B* ***64***, 060103 (2001).

22. Dupé, B. *et al.* Competing phases in BiFeO$_3$ thin films under compressive epitaxial strain. *Phys. Rev. B* **81**, 144128 (2010).
23. Lubk, A., Gemming S. & Spaldin N.A. First-principles study of ferroelectric domain walls in multiferroic bismuth ferrite. *Phys. Rev. B* **80**, 104110 (2009).
24. Ross, N. L., Zhao, J., and Angel, R. J., High-pressure single-crystal X-ray diffraction study of YAlO$_3$ perovskite. *J. Solid State Chem.* **177**, 1276 (2004).
25. Bennett, C. J. C. *et al.* Compositional tuning of the strain-induced iso-symmetric phase transition and of ferromagnetism in Bi$_{1-x}$Ba$_x$FeO$_{3-\delta}$. *Submitted.*
26. Rana, D. S. *et al.* Thickness dependence of the structure and magnetization of BiFeO$_3$ thin films on (LaAlO$_3$)$_{0.3}$(Sr$_2$AlTaO$_6$)$_{0.7}$ (001) substrate. *Phys. Rev. B* **75**, 060405 (2007).
27. Toupet, H., Le Marrec, F., Lichtensteiger, C., Dkhil, B. & Karkut, M. G. Evidence for a first-order transition from monoclinic *α* to monoclinic *β* phase in BiFeO$_3$ thin films. *Phys. Rev. B* **81**, 140101 (2010).
28. Haumont, R. *et al.* Effect of high pressure on multiferroic BiFeO$_3$. *Phys. Rev. B* **79**, 184110 (2009).
29. Lisenkov, S., Rahmedov, D. & Bellaiche, L. Electric-Field-Induced Paths in Multiferroic BiFeO$_3$ from Atomistic Simulations. *Phys. Rev. Lett.* **103**, 047204 (2009).
30. Noheda, B. and Cox D. E. Bridging phases at the morphotropic boundaries of lead oxide solid solutions. *Phase Transitions* **79**, 5 (2006).
31. Christen, H. M. & Eres, G. Recent advances in pulsed-laser deposition of complex-oxides. *J. Phys.: Condens.Matter* **20**, 264005 (2008).
32. Stokes, H. T. and Hatch, D. M. FINDSYM: program for identifying the space-group symmetry of a crystal. *J. Appl. Crystallogr.* **38**, 237 (2005); Stokes H. T., Hatch, D. M. & Campbell, B.J. (2007). ISOTROPY, stokes.byu.edu/isotropy.html.

**Acknowledgements**

H.S.K and H.M.C acknowledge support from the Laboratory Directed Research and Development Program of Oak Ridge National Laboratory, managed by UT-Battelle, LLC, for the U. S. Department of Energy. J.H.N. was supported by the Republic of Korea, Ministry of Knowledge and Economy, Visiting Scientists Program, under IAN: 16B642601, with the U.S. Department of Energy. NAS and AJF were supported by NSF Awards No. DMR-0820404 and No. NIRT-0609377. Computational resources used include the SGI Altix Cobalt system and the TeraGrid Linux Cluster Mercury at the National Center for Supercomputing Applications under Grant No. DMR-0940420 and CNSI Computer Facilities at UC Santa Barbara under NSF Grant No. CHE-0321368

**Author contributions**

J.H.N. and H.S.K. synthesized the samples used in this study. J.H.N. and H.M.C. performed the x-ray characterization and data analysis. A.J.H. and N.A.S. carried out the first-principles calculations.

**Supplementary information:**

|  | shear orientation | β | a/b | tilt pattern |
|---|---|---|---|---|
| *Cc* ($M_A$) | [110] | 88.1 | 1.00 | $a^-b^-c^0$ |
| *Pm* ($M_C$) | [100] | 90.0 | 1.00 | $a^-b^0c^0$ |
| *P1* | [100] | 88.4 | 1.01 | $a^-b^-c^0$ |

**Supplementary Table 1**: Optimized structural parameters for competing phases *Cc*, *Pm* and *P1*.

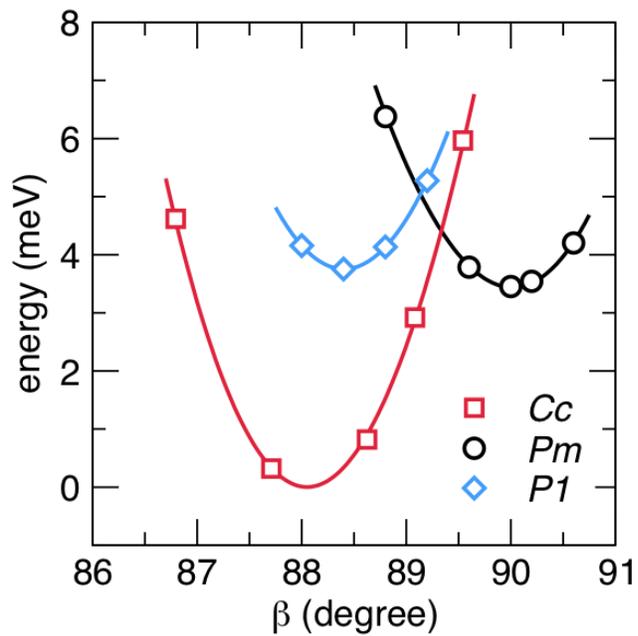

**Supplementary Figure 1:** Total calculated energy per formula unit of $BiFeO_3$ as a function of shear angle β. The optimized values of β are determined from a polynomial fit to each data set.

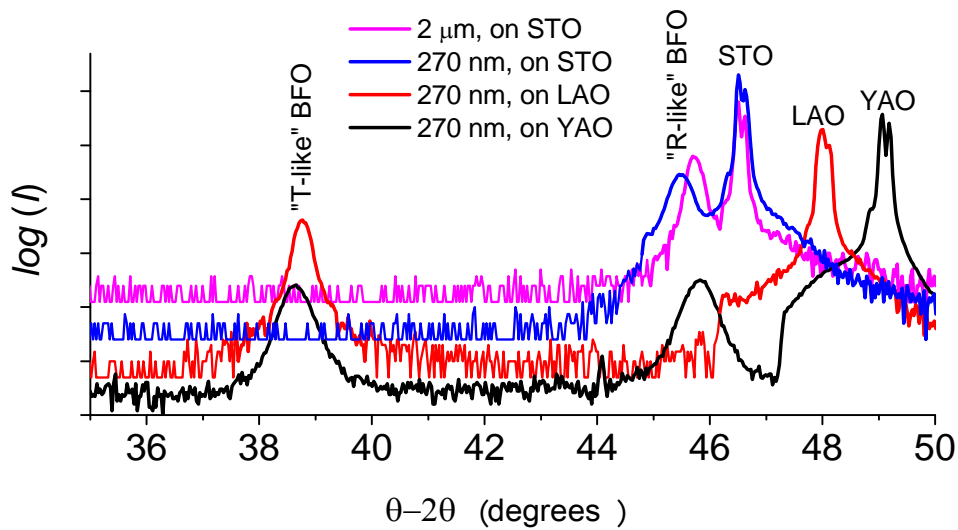

**Supplementary Figure 2:** Comparison of BiFeO$_3$ (BFO) films grown on different substrates. The semi-logarithmic plot shows results from X-ray θ-2θ scans for four BFO films of different thickness and on different substrates (offset vertically for clarity): SrTiO$_3$ (STO), LaAlO$_3$ (LAO), and YAlO$_3$ (YAO). The 002$_{pc}$ peaks for the film and substrates are shown. Films on SrTiO$_3$ are rhombohedral or "R-like", while the film on LaAlO$_3$ shows only the "T-like" phase. In contrast, the film on YAlO$_3$ shows a coexistence of both phases.